\documentclass{article}
\usepackage{amsmath}
\usepackage{graphicx}
\usepackage{float}
\usepackage{lineno}
\usepackage{authblk}
\usepackage{hyperref}
\usepackage[uniquelist=false, style=authoryear, maxnames=2]{biblatex}
 \usepackage[margin=1in]{geometry}
\usepackage{multirow}
\usepackage[T1]{fontenc}
\usepackage{newtxmath,newtxtext}

\title{A Bayesian Model to Estimate Abundance Based on Scarce Animal Vestige Data}

\author[1,2]{Niamh Mimnagh \thanks{Correspondence author. E-mail: niamh.mimnagh.2013@mumail.ie}}
\author[3]{Iuri Ferreira }
\author[4]{Luciano Verdade}
\author[1,2]{Rafael de Andrade Moral}
\affil[1]{\small Hamilton Institute, Maynooth University, Maynooth, Co. Kildare, Ireland}
\affil[2]{\small Department of Mathematics and Statistics, Maynooth University, Maynooth, Co. Kildare, Ireland}
\affil[3]{\small Federal University of S\~{a}o Carlos, S\~{a}o Carlos, Brazil }
\affil[4]{\small University of S\~{a}o Paulo, S\~{a}o Paulo, Brazil}
\date{}   
\begin{document}

\maketitle

\begin{abstract}
\begin{enumerate}
    \item We propose a modelling framework which allows for the estimation of abundances from trace counts. This indirect method of estimating abundance is attractive due to the relative affordability with which it may be carried out, and the reduction in possible risk posed to animals and humans when compared to direct methods for estimating animal abundance.
    \item We assess these methods by performing simulations which allow us to examine the accuracy of model estimates. The models are then fitted to several case studies to obtain abundance estimates for collared peccaries in Brazil, kit foxes in Arizona, red foxes in Italy and sika deer in Scotland.
    \item Simulation results reveal that these models produce accurate estimates of abundance at a range of sample sizes. In particular, this modelling framework produces accurate estimates when data is very scarce.
    \item The use of vestige counts in estimating abundance allows for the monitoring of species which may otherwise go undetected due to their reclusive nature. Additionally, the efficacy of these models when data is collected at very few transects will allow for the use of small-scale data collection programmes which may be carried out at reduced cost, when compared to larger-scale data collection. 
\end{enumerate}
\end{abstract}


\section{Introduction}
Estimating wildlife abundance may be relatively expensive and time-consuming. Therefore, whenever possible monitoring population fluctuation by an abundance index tends to be more cost-effective \parencite{nichols2014role}. However, in order to understand evolutionary-ecological processes and make decisions concerning wildlife management (i.e., conservation, use, coexistence, and monitoring, according to \cite{caughley1994directions}), one might need to know actual species abundance \parencite{verdade2014applied}. In addition, methods that involve capturing or even direct sightings of animals can be invasive and pose risks to both wildlife and humans \parencite{verdade2013counting}.
In this paper, we propose a modelling framework based on a triple Poisson hierarchy, which allows for the estimation of animal abundance from animal vestige count data, where a vestige may include any trace that an animal leaves behind as it moves through a study area. Here we examine data collected on animal droppings. However, this model may be adapted in the future for other traces which are not produced at a fixed rate, such as footprints, fur, or feathers. 

Advantages of the use of vestige data to estimate animal abundance include the reduced cost and labour required to carry out the survey, as well as the decreased disturbance caused to the animal, when compared with direct methods of estimating animal abundance \parencite{verdade2014applied}. In addition, we show that the modelling framework proposed here is especially useful to estimate animal abundance when data is scarce.

In Section \ref{methods} we introduce the modelling framework to estimate animal abundance from vestige count data. In Section \ref{simulation} we present simulation studies, which were carried out to assess the estimates of abundance when data is scarce, and to compare these estimates to those obtained using a distance sampling model. Finally, in Section \ref{cases} we present a number of case studies which we use to illustrate our modelling approach. 

\subsection{Related Works}
Several methods have been previously developed for estimating animal relative abundance based on vestige count data.

Distance sampling \parencite{Thomas2006, thomas2010distance} is a method which may be used for estimating animal density using vestige data, which involves modelling the assumption that the detectability of vestiges decreases with increasing distance from transects using a detection function, which can then be used to estimate animal density. \textcite{marques2001estimating} estimate the abundance of sika deer using this distance sampling methodology. In Section \ref{simulation} we provide a simulation study in which we compare abundance estimates obtained using a distance sampling model with those obtained using the novel triple Poisson model we propose.

\textcite{becker1991terrestrial, becker1998population, patterson2004estimating} propose a model for estimating abundance based on the observation of animal tracks in snow. This model assumes that the number of animal groups in the area may be obtained by following  tracks and locating  each group, while the model we propose here allows us to assign a prior to group size, which means we may estimate an unknown number of groups.

The Formozov–Malyshev–Pereleshin formula for estimating animal abundance is described by \textcite{stephens2006estimating}. This formula, originally proposed and published through Russian in \textcite{chelintsev1995mathematical}, involves estimating the probability that a transect will intersect an animal's track. This probability is then used to estimate the total number of track crossings, which can be used to estimate animal density. This formula requires both estimates of animal daily travel distances and counts of animal tracks whose age is known. 

\textcite{gallant2007unveiling, murray2005assessment, barnes2001reliable} examine the limitations associated with the use of scat surveys to estimate abundance. Among these limitations is the possibility of obtaining a false negative (a scenario in which animals are present at a certain site but do not leave any vestiges), the possibility for vestiges to be concentrated at certain sites, the possibility for vestiges produced at certain times of the year to decompose more quickly than those produced at other times, and the relatively small amount of the sample area that tends to be covered by transects. 

The triple Poisson model we propose requires the specification of a prior distribution on the number of groups in an area, the individuals per group, and the individual vestige production rate. Informative or non-informative priors may be provided in each case, depending on the level of prior knowledge available. This model may also be used in situations when data is scarce, and we provide examples which contain only two vestige counts to illustrate this.

\section{Methods}
\label{methods}
\subsection{Model Formulation}
This model assumes that we are examining closed populations of mammals which move around randomly in $G$ groups of size $N_i$, $i=1,\ldots,G$, within a study area which is homogeneous in terms of habitat use. We assume that
\begin{linenomath*}
\postdisplaypenalty=0
\begin{align*}
G\sim\mbox{Poisson}(\lambda_G),
\end{align*}
\end{linenomath*}
and therefore the total number of animals in the area is 
\begin{linenomath*}
\postdisplaypenalty=0
\begin{align*}
T=\displaystyle\sum_{i=1}^GN_i.
\end{align*}
\end{linenomath*}
By assuming the $N_i$ are independent and distributed as
\begin{linenomath*}
\postdisplaypenalty=0
\begin{align*}
N_i\sim\mbox{Poisson}(\lambda_N),
\end{align*}
\end{linenomath*}
we have that the conditional distribution of $T|G$ is \begin{linenomath*}
\postdisplaypenalty=0
\begin{align*}
T|G\sim\mbox{Poisson}(G\lambda_N).
\end{align*}
\end{linenomath*}

However, we do not observe realisations of $T$. Instead, we observe a fraction of the number of vestiges left by the animals. Let $V_t$ represent the total vestiges count at time $t$. We assume that the number of vestiges left at time $t=1$ has mean $\beta T$, i.e. depends on the individual vestige production rate $\beta$. By letting $V_1\sim\mbox{Poisson}(\beta T)$ and assuming vestiges left in the environment disappear exponentially over time with some constant rate, we may write \begin{linenomath*}
\postdisplaypenalty=0
\begin{align*}
V_t|V_1,\ldots,V_{t-1}\sim\mbox{Poisson}\left(\beta T+\displaystyle\sum_{j=1}^{t-1}\beta Te^{-\delta_{p}(t-j)}\right).
\end{align*}
\end{linenomath*}
where $\delta_{p}$ is a vestige decay parameter.

We are interested, however, in the limiting distribution of $V_t$, and for that we assume that after a short period of time, vestiges produced plus old vestiges that remain in the environment will be constant.
We next examine the rate term in the Poisson distribution above, which can be written as follows:  
\begin{linenomath*}
\postdisplaypenalty=0
\begin{align*}
   \beta T + \beta T \sum_{j=1}^{t}e^{-j \delta_{p}}
    &=\beta T e^{-(0)\delta_{p}}+\beta T \sum_{j=1}^{t}e^{-j \delta_{p}}=\beta T \sum_{j=0}^{t}e^{-j \delta_{p}}.
\end{align*}
\end{linenomath*}

\noindent We have that
\begin{linenomath*}
\postdisplaypenalty=0
\begin{align*}
\lim_{t\rightarrow\infty}\beta T \sum_{j=0}^{t}e^{-j \delta_{p}}=\displaystyle\frac{e^{\delta_{p} }\beta T}{e^{\delta_{p}}-1}.
\end{align*}
\end{linenomath*}

\noindent We can therefore let  $\alpha = \left(\frac{e^{\delta_{p}}\beta}{e^{\delta_{p}}-1}\right)$ and write the marginal distribution of $V=V_t$ as $V\sim\mbox{Poisson}(\alpha T)$.

By assuming random deposition of vestiges, the distribution of the observed number of vestiges will depend on the coverage level of the sampling method (e.g. transects, camera traps, etc). We refer to this coverage as $\nu\in(0,1)$. We assume that $\nu$ is known, and that every vestige within the covered area is detected.   Therefore, the distribution of the observed number of vestiges $Y$, alongside the full modelling hierarchy, can be written using the following triple Poisson hierarchical model:
\begin{linenomath*}
\postdisplaypenalty=0
\begin{align*}
Y|T,G,\nu &\sim \mbox{Poisson}(\alpha T \nu) \\
T|G  &\sim \mbox{Poisson}(G\lambda_N) \\
G &\sim \mbox{Poisson}(\lambda_G)
\end{align*}
\end{linenomath*}
The performance of the model is highly dependent on how well $\alpha$, the vestige production rate, is estimated. Therefore we may either opt to set up an informative prior for $\alpha$, or simply fix it as a ``known" value. To estimate $\lambda_G$ and $\lambda_N$ we may use informative or flat priors. However, the average group size is typically known, therefore we may use an informative prior for $\lambda_N$, and a flat gamma prior for $\lambda_G$.

It would also be reasonable to assume an aggregated process of resource allocation and/or aggregated animal behaviour. This would, in turn, affect the number of vestiges. A simple extension that would accommodate this assumption would be to treat the top tier of the hierarchy as an over-dispersed process, e.g. 
\begin{linenomath*}
\postdisplaypenalty=0
\begin{align*}
Y|T,G,\nu \sim \text{Negative Binomial}(\alpha T \nu, \phi)
\end{align*}
\end{linenomath*}
and use a flat gamma prior to estimate $\phi$, the over-dispersion parameter.

\section{Simulation Studies}
\label{simulation}
In this section we describe a simulation study to assess the accuracy of abundance estimates when data is scarce, and to compare estimates with those produced by a distance sampling model. Data was simulated from a distance sampling model, which required the specification of the size of the study area, the number of vestiges within the study area, the distance between transects, the truncation distance (the distance from a transect within which vestiges may be observed), and the detection function (which is used to model the distribution of vestiges given their distance from a transect). Data was simulated using the \texttt{DSsim} package \parencite{DSsim} using the R statistical software version 4.0.2 \parencite{RCT2020}, which saw 5000 vestiges contained within a $2 \times 5\text{km}$ area, with vestige density constant across the study area. A truncation distance  of  $10\text{m}$ was chosen, and transects were specified at a distance of $1000\text{m}$ from each other, which ensured that each simulation contained only two transects, each 5km long. As a result, the data used in the triple Poisson model comprises of just two numbers, i.e., the total number of vestiges observed at each transect. A half-normal detection function was used to model the probability of observing vestiges, given their distance from the transect, and the vestige density $D_{\text{vestige}}$ was subsequently obtained. This vestige density can now be used to determine animal abundance \parencite{marques2001estimating}. The total number of vestiges produced per day is calculated as $D_{\text{vestiges}}/\delta$,
where for the purpose of this simulation we let the true time to vestige decay $\delta$ be 10 days. The animal density $D$ can then be given as:
\begin{linenomath*}
\postdisplaypenalty=0
\begin{align*}
    D =\frac{\text{Total vestiges produced per day}}{\lambda},
\end{align*}
\end{linenomath*}
where the true individual vestige production rate $\lambda$ was allowed to be 15 per day. This density can then be used to calculate abundance as $N = D \times A$ where A is $10\text{km}^{2}$, the size of the study area.

We simulated 50 of these datasets. Triple Poisson models, with combinations of informative and non-informative $\lambda_{G}$ and $\lambda_{N}$ were fitted. The true abundance $N$ varied between datasets ($N \in \{25,45\}$). For this reason, in order to provide informative priors for $\lambda_{N}$ and $\lambda_{G}$, it was assumed that the mean number of animals per group might be between three and seven, and the mean number of groups in the area might be between one and ten. This allowed us to use the informative priors $\lambda_{N} \sim \text{Gamma}(5,1)$ and $\lambda_{G} \sim \text{Gamma}(10, 1)$.

The non-informative priors for $\lambda_{N}$ or $\lambda_{G}$ were assigned as a $\text{Gamma}(0.01,0.01)$. Data was simulated with 5000 vestiges present in the study area, so  the vestige production rate of the triple Poisson model $\alpha$ was given a weakly-informative Uniform$(10,10000)$ prior, where $\alpha$ represents the number of new vestiges produced per individual per day plus the number of vestiges still present in the area from previous days. These models were implemented using JAGS \parencite{plummer2003jags}, using the \texttt{R2jags} package \parencite{Su2020}.

Additionally, distance sampling models with combinations of $\delta$ and $\lambda$, specified correctly and incorrectly, were then fitted to this data using the \texttt{Distance} \parencite{JSSv089i01} package. The full list of models fitted are given in Table \ref{tab:models}. The accuracy of abundance estimates was assessed using relative bias for the true abundance $N$, averaged over all simulations, calculated as
\begin{linenomath*}
\postdisplaypenalty=0
\begin{align*}
\text{Relative mean bias} = \frac{\hat{N}-N}{N}.
\end{align*}
\end{linenomath*}
The smaller the value for relative bias, the closer to the true value our estimated abundances were.

\begin{table}[H]
    \centering
    \begin{tabular}{c c c c}
    \hline 
    \multicolumn{4}{c}{(a) Triple Poisson Models}\\
    Model & $\lambda_{N}$ & $\lambda_{G}$ & Abundance Relative Bias\\ 
    \hline
    TP1 & Informative  & Informative & 0.109\\
    TP2 & Informative & Non-Informative & 0.142\\
    TP3 & Non-Informative & Informative & 0.186 \\
    TP4 & Non-Informative & Non-Informative & 0.104\\
    &&\\
    \hline
    \multicolumn{4}{c}{(b) Distance Sampling Models}\\
    Model & $\lambda$ & $\delta$ & Abundance Relative Bias\\
    \hline
    DS1 &15&10 & 0.000\\
    DS2 & 16&10 & 0.091\\
    DS3 &15& 11 & 0.062\\
    DS4 &16& 11 & 0.147\\
    DS5 &17& 12 & 0.265\\
    DS6 &18& 13 & 0.359\\
    DS7 & 19 & 14 & 0.436\\
    \hline
    \end{tabular}
    \caption{(a) Triple Poisson models fitted with informative and non-informative Gamma priors on mean group size $\lambda_{N}$ and mean number of groups $\lambda_{G}$  (b) Distance Sampling models with individual vestige production ($\lambda$) and time to vestige decay ($\delta$) supplied correctly ($\lambda = 15,  \delta = 10$) and incorrectly, with incorrect values for $\lambda \in (16,19)$  and incorrect values for $\delta \in (11,14)$. .}
    \label{tab:models}
\end{table}

\begin{figure}[H]
    \centering
    \includegraphics[width=\textwidth]{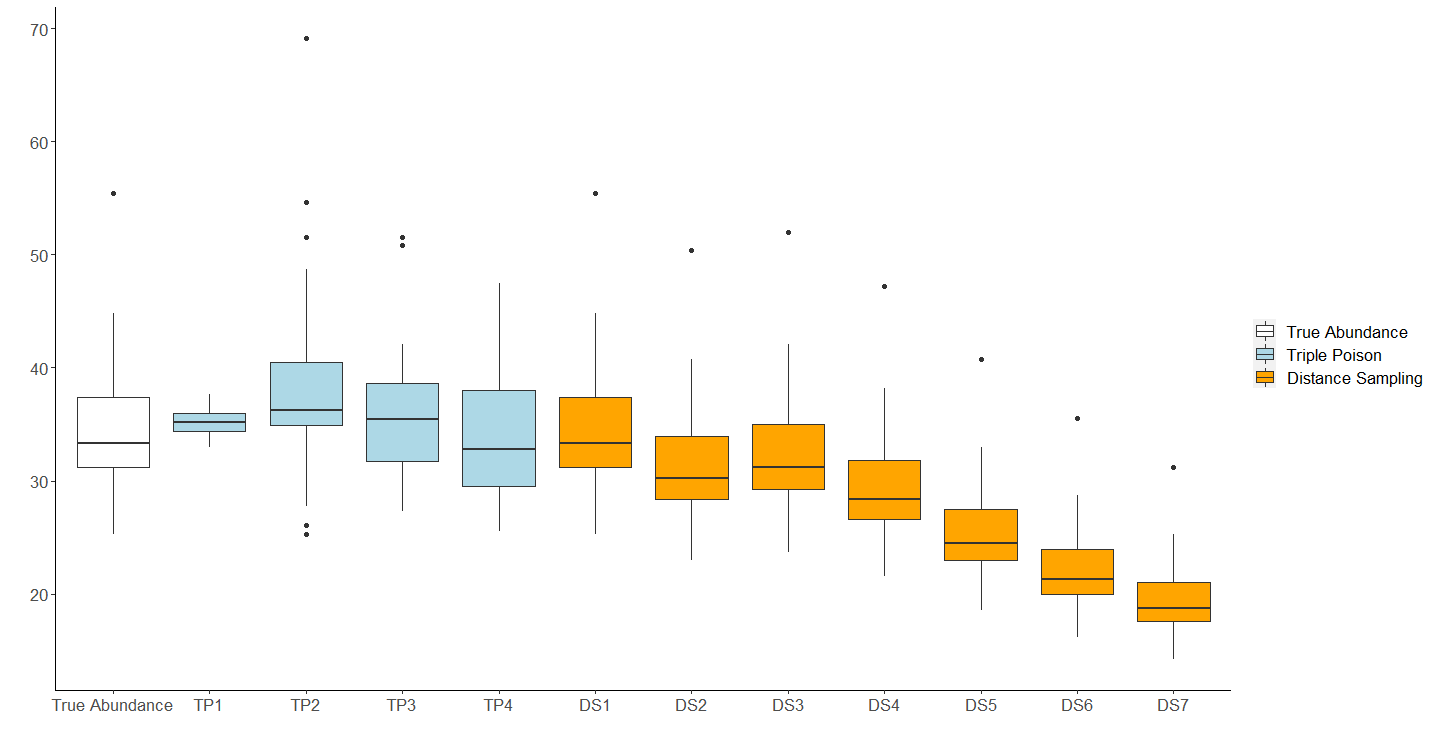}
    \caption{Boxplots comparing the true abundances with the abundances estimated from each of the triple Poisson (TP) and distance sampling (DS) models.}
    \label{fig:distance}
\end{figure}

\begin{figure}
     \centering
     \includegraphics[width=0.6\textwidth]{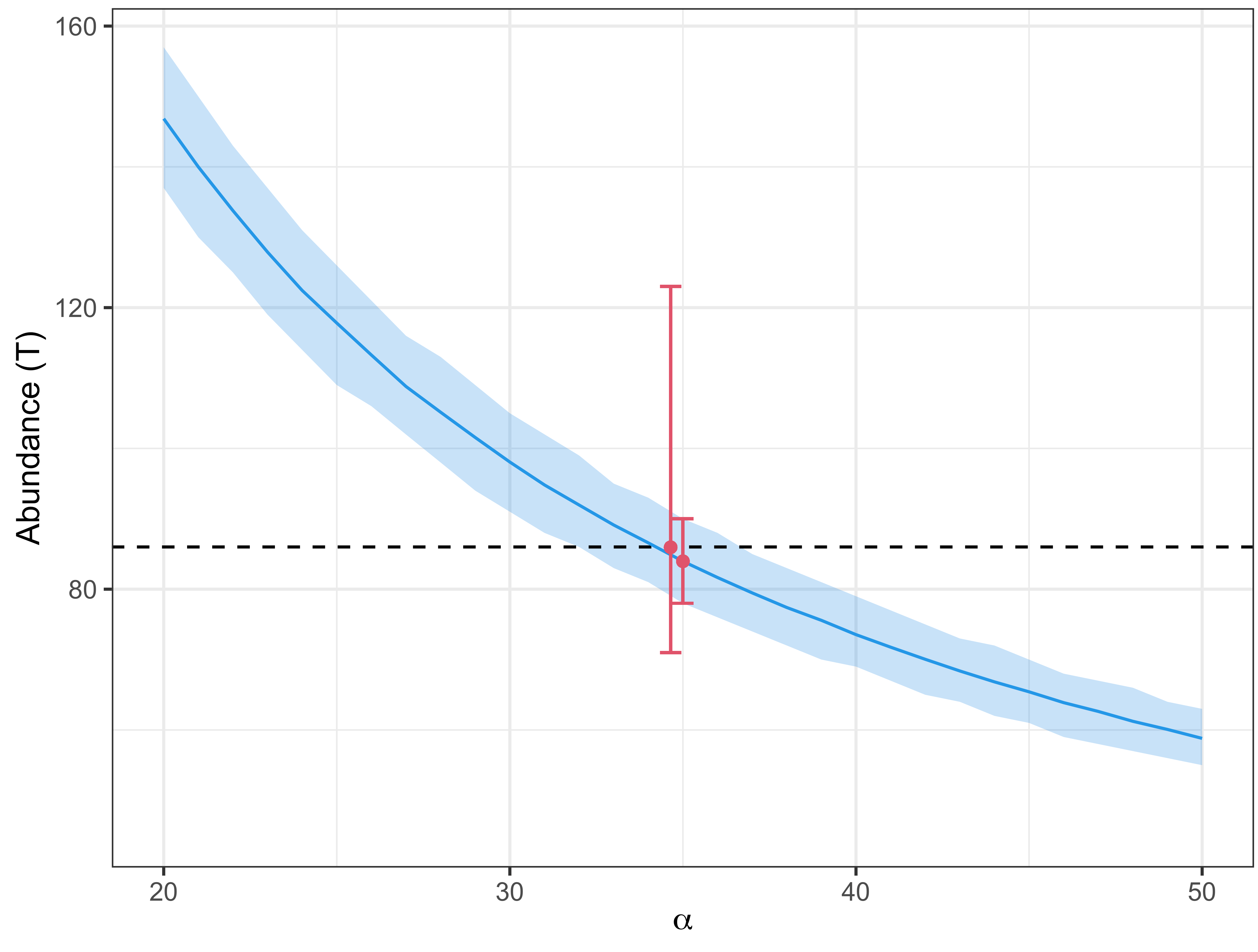}
     \caption{Abundance mean estimates and $95\%$ Credible intervals when the true value of $\alpha$ is known correctly (short crossbar), when $\alpha$ is known incorrectly (ribbon) and when $\alpha$ is supplied a Uniform prior (long crossbar).}
     \label{fig:alpha_simulation}
 \end{figure}

Simulations were run with the aim of determining the effect varying priors on $\alpha$ would have on abundance estimates. In this simulation there were three possible values for $\alpha$; one in which $\alpha$ is given a weakly-informative Uniform prior, one in which $\alpha$ is known correctly and supplied to the model as data, and one in which $\alpha$ is known incorrectly, and is supplied as data. The results are presented in Figure \ref{fig:alpha_simulation}, in which the true value for $\alpha$ is 35 and the true abundance is 86. The small crossbar represents the $95\%$ Credible Interval for the estimate of abundance when $\alpha$ is known correctly. From this we can see that when the true value for $\alpha$ is known, estimates of abundance are highly accurate and precise. The large crossbar represents the $95\%$ Credible interval for the estimate of abundance when a non-informative Uniform prior is supplied for $\alpha$. When $\alpha$ is unknown and is given a Uniform prior, the estimate for abundance has still got a high degree of accuracy, but is now imprecise. The ribbon represents the $95\%$ Credible Interval when an incorrect value is supplied for $\alpha$. In this case, the estimate for abundance is still very precise, but is now inaccurate. 
 
In addition to these scarce data simulations, extensive simulation studies were run to assess model estimation for different priors on $\lambda_{G}$, $\lambda_{N}$, and $\alpha$. Simulation studies were also run to determine the effects of a Poisson versus a Negative Binomial distribution on vestige count, and the effect on abundance estimates of temporally replicated vestige counts. The full details and results of all simulation studies can be found in Appendix A.

\section{Case Studies}
\label{cases}
The triple Poisson was fitted to data collected on different species at several locations. This model was first applied to these data sets assuming a Poisson conditional distribution for the observed vestiges $Y$. The models were subsequently run again, assuming a Negative Binomial conditional distribution for $Y$. 

\subsection{Collared Peccary}
\label{collaredPeccary}
The first case study is on the collared peccary (\textit{Dicotyles tajacu}), using vestige data collected in southeast Brazil \parencite{Assis2012}. Similar to our simulation studies, the data collected for this case study is comprised of vestige counts from only two transects. This means that the sample size available to us is very small, as it contains only two counts: $(7, 1)$.

However, prior information on group size and number of groups is available. Collared peccary groups  are estimated to be composed of between seven and nine individuals \parencite{Sowls1997}, and the number of groups in the area is estimated at between three and five. This allowed us to place informative priors on $\lambda_{G}$ and $\lambda_{N}$. However, information is not available on the individual vestige production rate, so we are unable to place an informative prior on $\alpha$ in this case. Here we use $\alpha \sim \text{Gamma}(0.01,0.01)$.

The transect along which seven vestiges were found was 8km long, while the other transect was 12km long. In order to estimate $v$, the transect coverage rate, we require the distance around the transect within which vestiges might be observed. This distance was not measured, so we make a conservative estimate that vestiges within 2m on either side of the transect are visible. This results in transects whose area is $32,000\text{m}^{2}$ and $48,000\text{m}^{2}$ respectively. The study area is $43.65\text{km}^{2}$, and we can therefore estimate the transect coverage rate as $v \in \{0.00073, 0.0011\}$.

Our model was fitted with first a Poisson distribution on the vestige count, and then a Negative Binomial distribution. The results were compared using DIC values, and the model with the lower DIC value was the one with a Negative Binomial distribution on the vestige count. The result obtained from this model was a mean estimate of 44 collared peccaries, with a $95\%$ Credible Interval of $(16,87)$.

\subsection{Kit Foxes}
A second case study examined is related to kit fox (\textit{Vulpes macrotis}) populations, using vestige data collected in the US State of Arizona \parencite{dempsey2014finding, dempsey2015evaluation}. This dataset contains vestige counts collected during three biological seasons (breeding, pup-rearing, and dispersal) between 2010 and 2013. The 2010-2012 data is collected at nine transects, each of which are 5km long. In 2013 an additional 38 transects of length 0.5km are added.

In this case, we also have information on kit foxes' habits that helps to inform our choice of prior for $\lambda_{N}$ and $\lambda_{G}$. Kit foxes typically live in pairs or small family groups, and have one to seven pups each year \parencite{morrell1971life}.
The study area is $879\text{km}^{2}$, and a kit fox's territory is in the range of $2.5-11\text{km}^{2}$. This allows us to estimate that up to 350 groups may be present within the study area. Foxes produce up to eight traces per day \parencite{WEBBON2004}. Initially transects were cleared of traces and 14 days later the survey was carried out, so we know that traces detected can only have been present for a maximum of 14 days. This allows us to place a weakly-informative prior on $\alpha$, i.e. $\alpha \sim \text{Uniform}(0,112)$.

Models were fitted comparing the use of a Poisson distribution on vestige count to a Negative Binomial distribution. The model with the lower DIC value in this case was the model with a Negative Binomial distribution on the vestige count. The result of this model fitting is a mean estimate of between 450 and 650 kit foxes within the study area between 2010 and 2013. Details of this model-fitting and results can be found in Appendix B.

\subsection{Sika Deer}
\label{sikadeer}
The data described in \textcite{marques2001estimating} is available as part of the \texttt{Distance} R package \parencite{JSSv089i01}. This data contains counts of sika deer (\textit{Cervus nippon}) droppings from eight regions in Scotland, as well as their distance from the transect, measured in centimetres. 
We use the \texttt{Distance} package \parencite{JSSv089i01} to obtain estimates of deer abundance using a distance sampling model, and then implement a triple Poisson model on the same data, which allows us to compare estimates provided by the two methodologies.

Sika deer live in groups of up to ten animals, which allows us to give an informative prior to $\lambda_{N}$. The territory of a sika deer is in the range of $0.02-0.12\text{km}^{2}$. Vestige production for sika deer is between 10 and 30 each day \parencite{marques2001estimating}, and traces may take several months to decompose, so we placed a weakly informative prior on $\alpha$, i.e. $\alpha \sim \text{Uniform}(0,3500)$.

As this data contains the distance between traces and the transect, we know that the largest distance at which traces were observed from the transect was two metres. For this reason, the triple Poisson models are run assuming that traces are visible within two metres on either side of the transect.

Each of the eight regions has a different area, so we must provide a different prior for $\lambda_{G}$ for each area. We also fitted each model with both a Poisson distribution and a Negative Binomial distribution on the vestige count. The details of model implementation per area are available in Appendix B. A comparison of the abundance estimates obtained from the triple Poisson and distance sampling models can be found in Table \ref{tab:sika_distance_compare}.

\begin{table}[H]
    \centering
    \begin{tabular}{c rr rr}
    \hline
   \multirow{2}{*}{Area} &   \multicolumn{2}{c}{Distance Sampling}   &  \multicolumn{2}{c}{Triple Poisson}\\
   & \multicolumn{1}{c}{Estimate} & \multicolumn{1}{c}{95\% CI} & \multicolumn{1}{c}{Estimate} & \multicolumn{1}{c}{95\% CI} \\
   \hline
    A & 1027 & (690, 1528) & 1089 & (631, 2124)\\
    B & 382 & (219, 667) & 385 & (174,887)\\
    C & 33 & (15, 74) & 83 & (21, 316)\\
    E & 29 & (8, 99) & 90 & (19,334)\\
    F & 209 & (173, 252) & 317 & (141, 848)\\
    G & 125 & (18, 856) & 281 & (84, 786) \\
    H & 17 & (14, 21) & 83 & (11, 342)\\
    J & 69 & (57, 83) & 151 & (38, 526)\\
    \hline
    \end{tabular}
    \caption{Mean estimates and their 95\% confidence (distance sampling) or credible (Triple Poisson) intervals for sika deer abundance per area from a distance sampling model and the triple Poisson model.}
    \label{tab:sika_distance_compare}
\end{table}

\subsection{Red Foxes}
\textcite{cavallini1994faeces} collected data on red fox (\textit{Vulpes vulpes}) vestiges from nine transects in central Italy, which allowed them to obtain an index of fox abundance, estimated as the number of vestiges per km. This data was collected once every month for a year, so it is a good candidate for our modelling framework which takes into account temporal replicates. 

We have prior knowledge on common group sizes for red foxes, as well as their territory size, which allows us to place informative priors on $\lambda_{G}$ and $\lambda_{N}$. Red foxes have as many as nine cubs in a litter, which means a group may contain as many as 11 foxes.  Their territory ranges from $5$ to 12km$^{2}$. In a 2448km$^{2}$ area there could be as many as 490 groups. Transect coverage in this case was again calculated using transect length and study area, assuming that vestiges within 2m of the transect are visible. As with kit foxes, red foxes may produce eight traces per day. However, this data was collected monthly over a 12 month period, so we remove the initial observations (April 1992) for each region, as these traces may have been present for several months, and the longest time that the remaining traces may have been present in the environment was approximately 30 days. The prior used for $\alpha$ in this case was $\alpha \sim \text{Uniform}(0,240)$.

Our model was fitted with first a Poisson distribution on the vestige count, and then a Negative Binomial distribution. The results were compared using DIC values, and the model with the lower DIC value was the one with a Negative Binomial distribution on the vestige count. The result obtained from this model was a mean estimate of 3304 red foxes, with a $95\%$ Credible Interval of $(2603,4442)$. In 1994 Cavallini was able to obtain an index of abundance per region, calculated as the number of traces per km of transect, but was not able to produce abundance estimates as we do with the triple Poisson model.

\section{Discussion}

In this study, data was simulated from a distance sampling model, and distance sampling and triple Poisson models were fitted to compare abundance estimates. While distance sampling models consistently produce accurate estimates of abundance when vestige production rate $\lambda$ and time to vestige decay $\delta$ are known precisely (Table \ref{tab:models} (b)), small inaccuracies in $\lambda$ and $\delta$ produce estimates of abundance with relative bias that increases quite rapidly. Triple Poisson models were fitted with priors on $\lambda_{N}$ and $\lambda_{G}$ informative and non-informative. The result was estimates of abundance with relative bias $\in (0.1,0.2)$ regardless of prior choice for $\lambda_{N}$ or $\lambda_{G}$. In Figure \ref{fig:distance}, the triple Poisson model with non-informative priors on $\lambda_{G}$ and $\lambda_{N}$ better captures the distribution of the actual abundances, while estimates from models with two informative priors are clustered around the median.
Distance sampling models appear quite sensitive to changes in $\lambda$ and $\delta$ while the triple Poisson model seems more robust to changes in informativity in priors on $\lambda_{G}$ and $\lambda_{N}$. These scarce data simulation studies are representative of the type of data available in Section \ref{collaredPeccary}, wherein we have just one vestige count collected at each of two transects. In Section \ref{sikadeer} we compare the use of a distance sampling model and the triple Poisson model using scat survey data. 

Other simulations were performed to determine the accuracy of abundance estimates in various situations (details of these simulations are available in Appendix A). Results of these simulations reveal that the triple Poisson model suffers from identifiability issues in regards to $\lambda_{G}$ and $\lambda_{N}$ when the priors for these parameters are not informative. However, despite these identifiability issues, the abundance is still estimated well. Simulations also reveal that vestige production rate $\alpha$ is confounded with abundance, which is expected given the model formulation. When $\alpha$ is a known value, estimates for abundance are not dependent on priors supplied for $\lambda_{G}$ and $\lambda_{N}$. However when $\alpha$ is unknown and must be estimated from the model, abundance estimates are dependent on priors for $\lambda_{G}$ and $\lambda_{N}$, with the most accurate estimate abundances produced when these priors are informative.

While a weakly-informative Uniform prior is used in most cases here, which allows for the incorporation of prior information on vestige production rates per animal, the triple Poisson model may still be used if this information is not available. As in the collared peccary case study, if we do not have any prior knowledge to inform $\alpha$ we may instead use a non-informative Gamma prior. The result of this is abundance estimates which, while not as precise (indicated by wider credible intervals), will still produce usable mean abundance estimates.

In this study we assumed vestiges experience an exponential rate of decay. However, in the future we will examine alternative scenarios, in which vestiges may decay according to different distributions, to allow us to model the transient phase of the population. 

In the future, we also plan to examine the effect of convenience sampling rather than the use of randomly placed transects on the estimation of abundance. This might facilitate field work on data collection and improve cost-efficiency of animal counting.
The decision-making process concerning wildlife management is usually based on the cost-efficiency relations of survey/monitoring methods available \parencite{nichols2014role}. Traditional methods involving capturing and/or direct sight seen of animals tend to be invasive, time-consuming, and relatively expensive \parencite{verdade2013counting, verdade2014applied}. In addition, traditional methods of animal counting tend to have low precision and unknown accuracy \parencite{verdade2014applied}. The use of vestiges on the estimation of actual abundance or population density simplifies the process of animal counting; therefore improving the decision-making process concerning wildlife management.

\section{Acknowledgements}
Niamh Mimnagh's work was supported by a Science Foundation Ireland grant number 18/CRT/6049. The opinions, findings and conclusions or recommendations expressed in this material are those of the author(s) and do not necessarily reflect the views of the Science Foundation Ireland.

Luciano Verdade's work is supported by a Fundação de Amparo à Pesquisa do Estado de São Paulo (FAPESP) grant (FAPESP Proc. No. 2017/01304-4).

\section{Data Accessibility}
Collared peccary data is provided in its entirety in Section \ref{collaredPeccary}. 

\noindent Kit fox data is available as supplementary material in Dempsey et al., 2014 (accessed via \url{https://www.ncbi.nlm.nih.gov/pmc/articles/PMC4605691/} on 11/04/2022)

\noindent Sika deer data is accessible through the distance R package \parencite{JSSv089i01}.

\noindent A table containing the red fox data can be found in Cavaillini's 1994 paper. 

\section{Author's Contributions}
All authors contributed to methodology design. NM analysed the data and carried out the simulation studies. NM and RAM led the writing of the manuscript. All authors contributed critically to the drafts and gave final approval for publication.

\printbibliography
\newpage
\appendix
\section{Simulation Studies}
In this section we present simulation studies, which were run to assess the accuracy of abundance estimates. A range of scenarios were considered. Both Poisson and negative binomial distributions on vestige count $Y$ were evaluated, and in the case of the negative binomial model an extra parameter is introduced; the overdispersion parameter, $\phi \in \{0.2,2\}$, where the smaller the value of $\phi$, the greater the degree of overdispersion in the data. We then examined abundance estimates when vestige production rate $\alpha$ is assumed known and provided as data, and compared this to when $\alpha$ is unknown and must be estimated from the model using a Uniform distribution. The true value for $\alpha$ was 20, and $\alpha$ was estimated using a Uniform(0, 50) distribution. We next considered the use of both informative and non-informative Gamma priors on group size $\lambda_{G}$ and number of groups $\lambda_{N}$. True values were $\lambda_{N} \in \{5,10\}$ and $\lambda_{G} \in \{5,10\}$ so a non-informative Gamma prior was a Gamma(0.01,0.01) and an informative Gamma prior was Gamma(3,1), the latter of which has a mean of 3, and so is expected to provide a good fit when $\lambda_{N}$ and $\lambda_{G}$ are equal to five, but a poor fit when $\lambda_{N}$ and $\lambda_{G}$ are equal to 10. Additionally we examined a Uniform(1,100) prior on both group size and number of groups, which has a mean of approximately 50 and so is expected to provide a poor fit in all scenarios. The effect of sample size was assessed by comparing results for a single count collected at 10 sites, a single count collected at 100 sites, and 10 counts collected at 10 sites.  The result was a simulation study which contained 288 total variable combinations, each of which was simulated 100 times. In each case we estimated $\lambda_{N}$, $\lambda_{G}$, the total number of individuals $T$ and the vestige production rate $\alpha$.

\begin{figure}[H]
    \centering
    \includegraphics[width=0.9\linewidth]{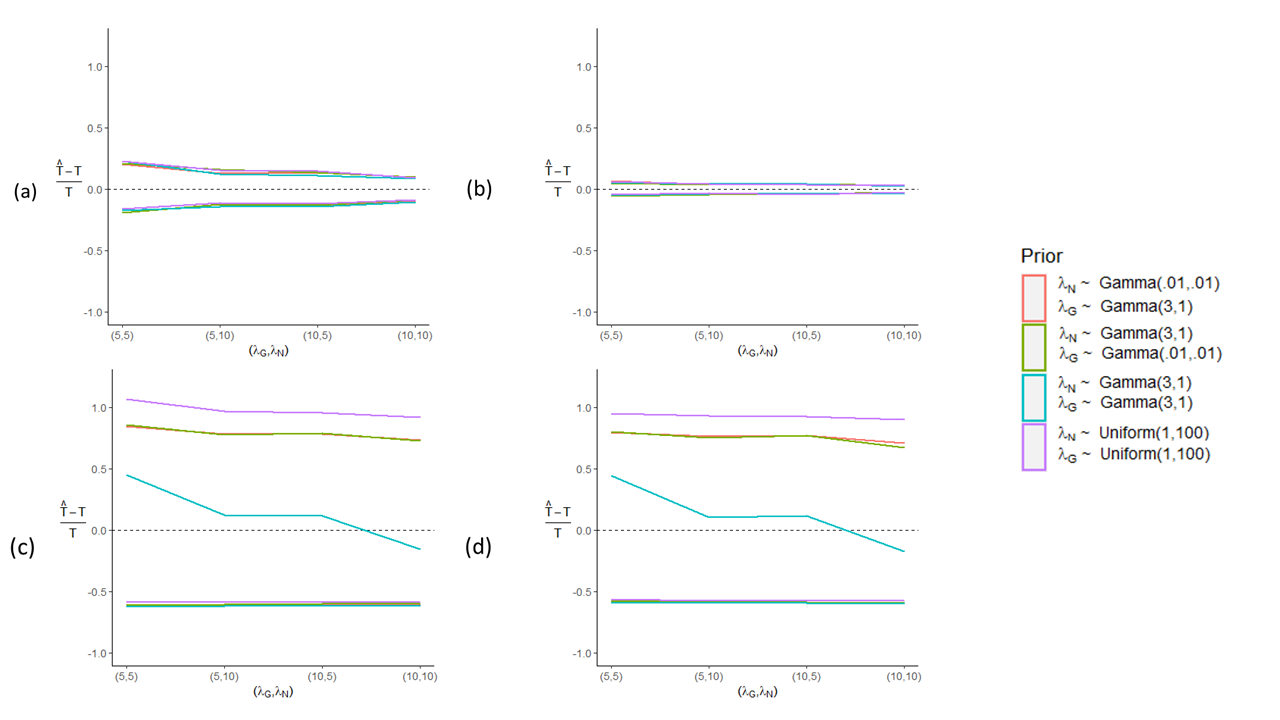}
    \caption{Mean relative bias in abundance $95\%$ credible intervals, obtained using  informative and non-informative priors for $\lambda_{G}$ and $\lambda_{N}$ for four scenarios in which vestige counts follow a Poisson distribution and a single count is collected per transect: (a) data collected at 10 transects with  $\alpha$ correctly specified (b) data collected at 100 transects with $\alpha$ correctly specified (c) data collected at 10 transects with $\alpha$ estimated by a Uniform prior(d) data collected at 100 transects with $\alpha$ estimated by a Uniform prior.}
    \label{fig:single_poisson_sim}
\end{figure}

\begin{figure}[H]
    \centering
    \includegraphics[width=0.9\linewidth]{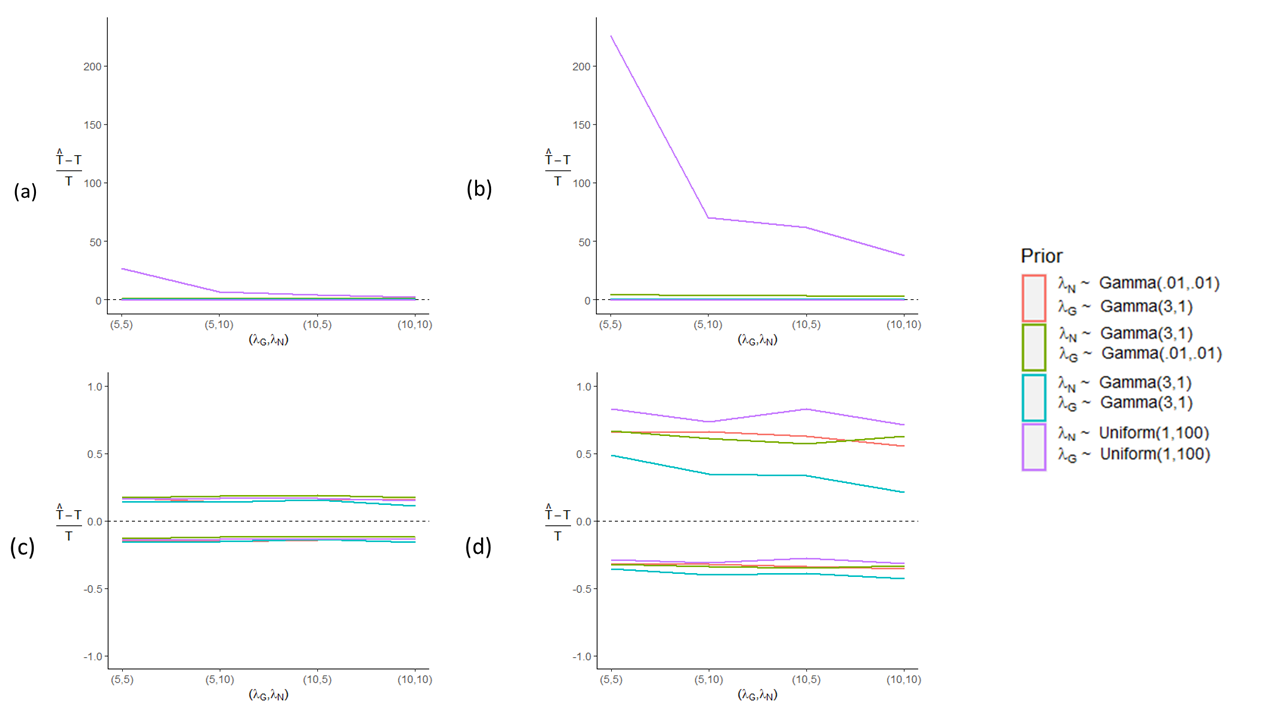}
    \caption{(a) correct $\alpha$, 10 transects, small overdispersion (b) correct $\alpha$, 10 transects, large overdispersion, (c) correct $\alpha$, 100 transects, small overdispersion, (d) correct $\alpha$, 100 transects, large overdispersion}
    \label{fig:single_negbin_correct_alpha}
\end{figure}

\begin{figure}[H]
    \centering
    \includegraphics[width=0.9\linewidth]{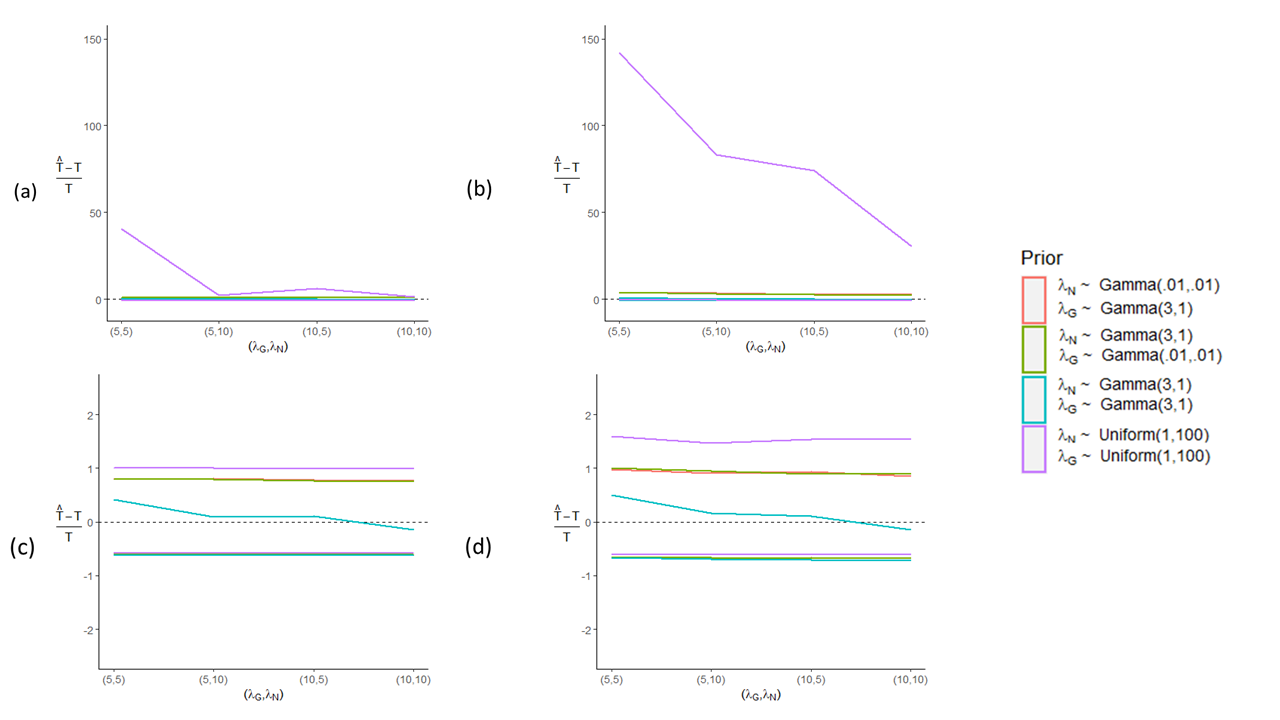}
    \caption{(a) Uniform $\alpha$, 10 transects, small overdispersion (b) Uniform $\alpha$, 10 transects, large overdispersion, (c) Uniform $\alpha$, 100 transects, small overdispersion, (d) Uniform $\alpha$, 100 transects, large overdispersion}
    \label{fig:single_negbin_uniform_alpha}
\end{figure}

\begin{figure}[H]
    \centering
    \includegraphics[width=0.9\linewidth]{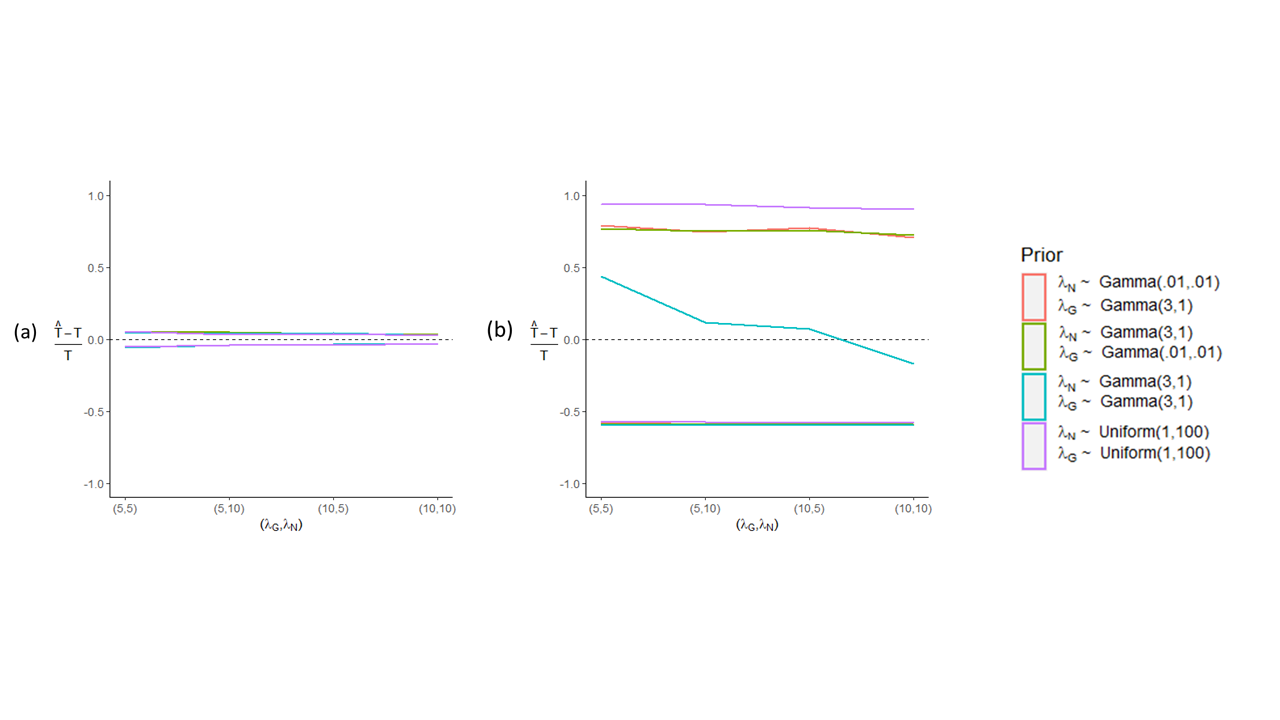}
    \caption{Temporally replicated counts, each with Poisson Y and 10 transects (a) Correct $\alpha$, (b) Uniform $\alpha$}
    \label{fig:temporal_replicate_poisson}
\end{figure}

\begin{figure}[H]
    \centering
    \includegraphics[width=0.9\linewidth]{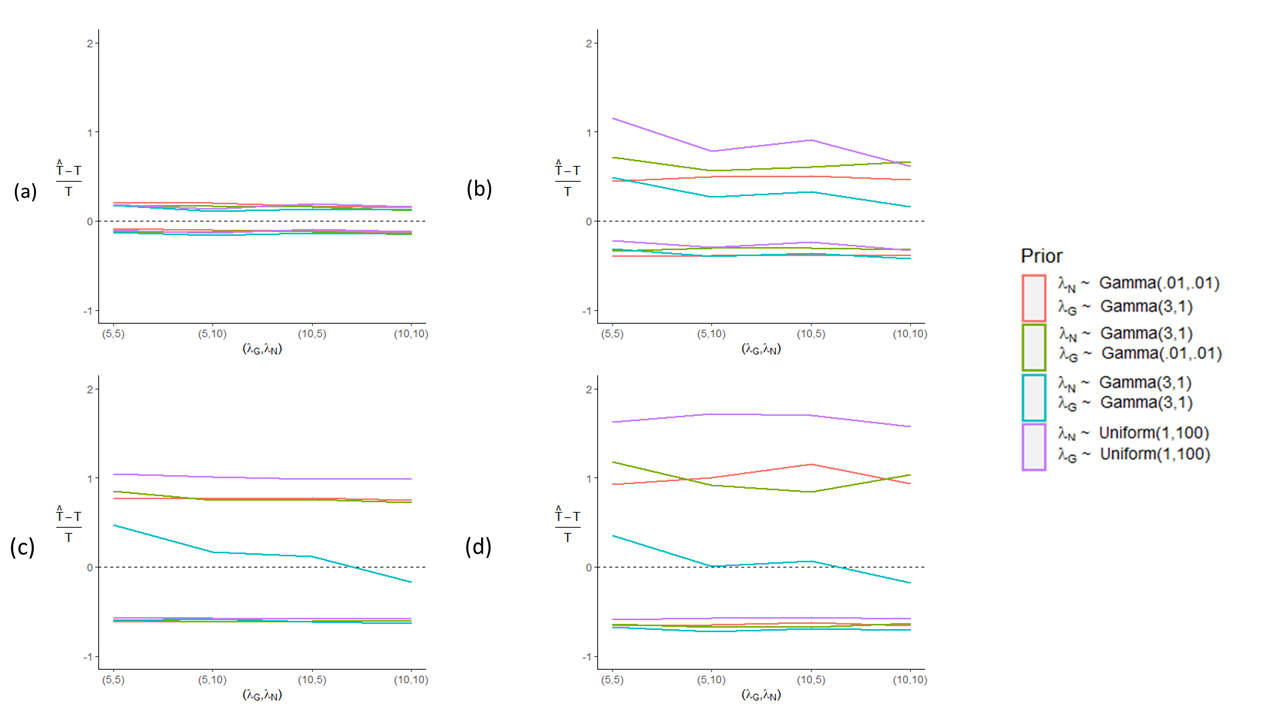}
    \caption{Temporally replicated counts, each with Negative Binomial Y and 10 transects (a) correct $\alpha$, small overdispersion, (b) correct $\alpha$, large overdispersion, (c) Uniform $\alpha$, small overdispersion, (d) Uniform $\alpha$, large overdispersion}
    \label{fig:temporal_replicate_negbin}
\end{figure}

Figures \ref{fig:single_poisson_sim} to \ref{fig:temporal_replicate_negbin} each present mean relative bias in credible intervals coverage for the abundance estimates, calculated as $\frac{\hat{T}_{\text{upper}}-T}{T}$ and $\frac{\hat{T}_{\text{lower}}-T}{T}$, and averaged across 100 simulations, where $\hat{T}_{\text{upper}}$ and $\hat{T}_{\text{lower}}$ are the upper and lower bounds of the $95\%$ credible interval, respectively. Each figure is composed of two to four panels, with true values for $\lambda_{N}$ and $\lambda_{G}$ on the x-axis, and relative bias in abundance estimates on the y-axis.  Each panel displays relative bias in abundance for four different choices of prior on $\lambda_{N}$ and $\lambda_{G}$.

In Figure \ref{fig:single_poisson_sim} we present results for a simulation in which abundance counts follow a Poisson distribution, and a single sample is collected per transect. Figure \ref{fig:single_poisson_sim}(a) and Figure \ref{fig:single_poisson_sim}(c) contain data collected from 10 transects, while Figure \ref{fig:single_poisson_sim}(b) and Figure \ref{fig:single_poisson_sim}(d) contains data from 100 transects. Figure \ref{fig:single_poisson_sim}(a) and Figure \ref{fig:single_poisson_sim}(b) have $\alpha$ supplied as data, while Figure \ref{fig:single_poisson_sim}(c) and Figure \ref{fig:single_poisson_sim}(d) have $\alpha$ estimated using a Uniform distribution. When $\alpha$ is supplied as data, the choice of prior has little effect on the relative bias of T. In this case, the larger sample size provided by 100 transects results in lower relative bias when compared to 10 transects.
When $\alpha$ is estimated from the data the situation is very different, however. Now an informative prior on $\lambda_{N}$ and $\lambda_{G}$ is required to reduce relative bias, though an ill-fitting informative prior will produce inaccurate results, as is evident when $\lambda_{N} = \lambda_{G} = 10$.

Figure \ref{fig:single_negbin_correct_alpha} sees $\alpha$ correctly specified as data. In Figure \ref{fig:single_negbin_correct_alpha}(a) and Figure \ref{fig:single_negbin_correct_alpha}(b) (which each have data collected from 10 transects) the ill-fitting Uniform distribution provides abundance estimates with very high relative bias. This issue is exasperated when the degree of overdispersion in the data is large (Figure \ref{fig:single_negbin_correct_alpha}(b)), and is somewhat resolved by the larger sample size of 100 transects (Figure \ref{fig:single_negbin_correct_alpha}(c), which has small overdispersion, $\phi=2$, and Figure \ref{fig:single_negbin_correct_alpha}(d) with large overdispersion, $\phi=0.2$). Relative bias in the estimate of abundance is much smaller at the large sample size, with the smallest relative bias associated with the scenario when overdispersion is small and sample size is large. 

In Figure \ref{fig:single_negbin_uniform_alpha} we present the results when vestige count follows a Negative Binomial distribution, $\alpha$ is estimated using a Uniform distribution, and a single sample is collected per transect. Results are somewhat similar to those presented in Figure \ref{fig:single_negbin_correct_alpha}, though in all cases, relative bias is higher in Figure \ref{fig:single_negbin_uniform_alpha}. Estimating $\alpha$ rather than providing it as data results in greater bias in abundance estimates. In Figure \ref{fig:single_negbin_uniform_alpha}, estimating $\alpha$ results in relative bias estimates which are more dependent on the choice of prior for $\lambda_{N}$ and $\lambda_{G}$, and the misinformed Gamma(3,1) prior provides biased estimates when $\lambda_{N}=\lambda_{G}=10$.

In Figure \ref{fig:temporal_replicate_poisson}, we present results when vestige count uses a Poisson distribution, and each of 10 transects is sampled on 10 occasions.  Results produced from this simulation are very similar to those presented in Figure \ref{fig:single_poisson_sim}(b) and Figure \ref{fig:single_poisson_sim} (d) -- using 10 transects with data collected at 10 time points produces results equivalent to those produced when 100 transects are each surveyed only once. When $\alpha$ is supplied as data (Figure \ref{fig:temporal_replicate_poisson}(a)), abundance is estimated with high levels of accuracy, and these estimates do not depend on the size of $\lambda_{G}$ and $\lambda_{N}$ or the priors supplied for them. When $\alpha$ is estimated (Figure \ref{fig:temporal_replicate_poisson}(b)) estimates of abundance are dependent on the priors supplied for $\lambda_{G}$ and $\lambda_{N}$, with most accurate estimates produced when both $\lambda_{G}$ and $\lambda_{N}$ have informative (but not misinformed, as we see when a Gamma(3,1) prior is used and $\lambda_{N}=\lambda_{G}=10$) priors.

In Figure \ref{fig:temporal_replicate_negbin}, we present results when vestige count has a Negative Binomial distribution and 10 counts are collected for each of 10 transects. These results  appear similar to those produced by a Negative Binomial model with 100 transects (Figure \ref{fig:single_negbin_correct_alpha} and Figure \ref{fig:single_negbin_uniform_alpha}). When $\alpha$ is specified correctly the prior for $\lambda_{N}$ and $\lambda_{G}$ do not have a large impact on relative bias for abundance, particularly when overdispersion is small. However, when $\alpha$ is estimated with a Uniform distribution, the prior on $\lambda_{G}$ and $\lambda_{N}$ has a greater effect on abundance estimates, with relative bias slightly higher when overdispersion is large, and the misspecified Gamma prior producing biased estimates when mean group size and mean number of groups are both 10.

\section{Case Studies}
In this section we present the details of model implementation for each of our four case studies. 

\subsection{Sika Deer}
Each of the  regions in the sika deer dataset is of a different size. In order to determine the prior to use for  $\lambda_{G}$, we use the size of the region, and prior knowledge on the animal's habitat size. The territory of sika deer lies between two and twelve hectares. Region areas, the maximum number of groups that area could contain, and the prior used for $\lambda_{G}$ are given in Table \ref{tab:sika_deer_prior}.

\begin{table}[H]
    \centering
    \begin{tabular}{c c c c}
    \hline
       Region  &  Area & Max. Groups & Prior\\
       \hline
       A  & $13.9\text{km}^{2}$ & 700 & $\text{Gamma}(1,72)$\\
       B & $10.3\text{km}^{2}$ & 500 & $\text{Gamma}(1,45)$\\
       C &$8.6\text{km}^{2}$ & 430&$\text{Gamma}(1,40)$ \\
       E &$8\text{km}^{2}$ & 400&$ \text{Gamma}(1,40)$\\
       F & $14\text{km}^{2}$&700 & $ \text{Gamma}(1,72)$\\
       G &$15.2\text{km}^{2}$ & 760& $\text{Gamma}(1,75)$\\
       H &$11.3\text{km}^{2}$ & 565&$\text{Gamma}(1,52)$ \\
       J &$9.6\text{km}^{2}$ & 480& $\text{Gamma}(1,50)$\\
       \hline
    \end{tabular}
    \caption{The area of each region, as well as the maximum number of groups of sika deer which might inhabit this area and the prior used for $\lambda_{}G$}
    \label{tab:sika_deer_prior}
\end{table}

\subsection{Kit Foxes}
Given that the area is $879\text{km}^{2}$, and we have prior knowledge that a Kit fox's territory is in the range of $2.5-11\text{km}^{2}$, we estimate that up to 350 groups may be present within the study area. This allows us to place a prior on $\lambda_{G}\sim \text{Gamma}(1,45)$.

\begin{figure}[H]
    \centering
    \includegraphics[width=\textwidth]{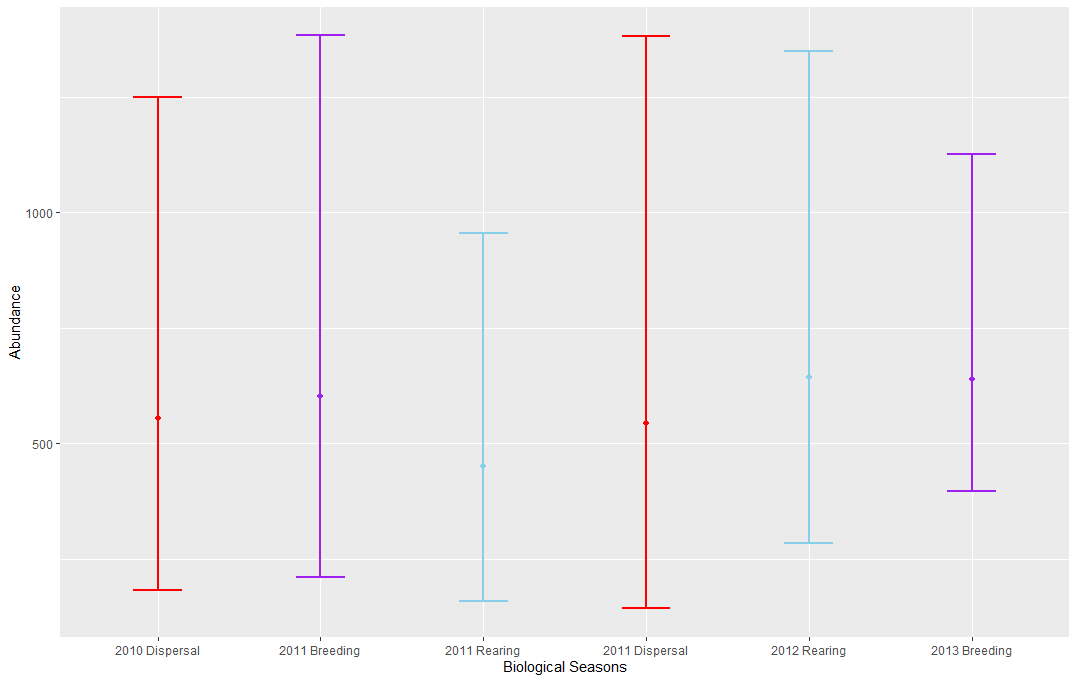}
    \caption{Abundance estimates at different biological seasons between 2010 and 2013.}
    \label{fig:kf_abundance}
\end{figure}

Kit fox data from six different biological seasons was analysed. The abundance estimates for each biological season are provided in Figure \ref{fig:kf_abundance}. There appears to be little change in mean abundance during this time, and most biological seasons have an abundance of 500-600 individuals.

\subsection{Red Foxes}
The territory of a red fox ranges from $5$km$^{2}$ to $12$km$^{2}$, and the size of the study area is $2448$km$^{2}$, so there may be as many as 490 groups in the area. For this reason, we let $\lambda_{G}\sim \text{Gamma}(1,50)$, which allows a maximum of between 400 and 500 groups.

\section{DIC Values}
Table \ref{tab:DIC} provides the DIC values for each case study. In each case, the model with the lower DIC value was chosen as the best fit.

\begin{table}[h]
    \centering
    \begin{tabular}{ccc}
    \hline
      Model   &  Poisson & Negative Binomial \\
      \hline
      Collared Peccary   & 16.12 & 15.14\\
      Kit Foxes & 259.25& 87.32\\
      Red Foxes & 1195.94 & 629.33\\
      Sika Deer A & 555.65 & 145.41\\
      Sika Deer B &283.99& 94.41\\
      Sika Deer C &14.92& 15.52\\
      Sika Deer E &46.17& 36.34\\
      Sika Deer F &7.34& 8.23\\
      Sika Deer G&34.83& 27.38\\
      Sika Deer H & 5.09& 5.42\\
      Sika Deer J &5.82 & 6.26\\
      \hline
    \end{tabular}
    \caption{DIC Values for models fitted to case studies in which we compare a Poisson vestige count to a Negative Binomial vestige count.}
    \label{tab:DIC}
    \end{table}
\end{document}